\documentclass[aps,prb,reprint,amssymb,amsmath,floatfix]{revtex4-1}

\usepackage[utf8]{inputenc}
\usepackage[T1]{fontenc}

\usepackage{times}
\usepackage{newtxmath}
\usepackage{bm}

\usepackage[english]{babel}
\usepackage{csquotes}

\usepackage{microtype}
\usepackage[colorlinks=true,allcolors=blue]{hyperref}
\usepackage{lipsum}

\usepackage{graphicx}

\usepackage{siunitx}
\usepackage{booktabs}

\usepackage{comment}
\usepackage{soul}
\usepackage{xcolor}
\definecolor{DarkRed}{rgb}{0.65,0,0}
\definecolor{DarkBlue}{rgb}{0,0,0.65}

\renewcommand{\vec}[1]{\bm{#1}}
\providecommand{\drm}{\ensuremath{\mathrm{d}}}
\DeclareMathOperator{\csch}{csch}

\begin{document}

\title{Universal Absence of Walker Breakdown and Linear Current--Velocity Relation \\ via Spin--Orbit Torques in Coupled and Single Domain Wall Motion}
\author{Vetle Risinggård}
\email{vetle.k.risinggard@ntnu.no}
\author{Jacob Linder}
\affiliation{Department of Physics, NTNU, Norwegian University of Science and Technology, N-7491 Trondheim, Norway}
\date{\today}

\begin{abstract}
We consider theoretically domain wall motion driven by spin--orbit and spin Hall torques. We find that it is possible to achieve universal absence of Walker breakdown for all spin--orbit torques using experimentally relevant spin--orbit coupling strengths. For spin--orbit torques other than the pure Rashba spin--orbit torque, this gives a linear current--velocity relation instead of a saturation of the velocity at high current densities. The effect is very robust and is found in both soft and hard magnetic materials, as well as in the presence of the Dzyaloshinskii--Moriya interaction and in coupled domain walls in synthetic antiferromagnets, where it leads to very high domain wall velocities. Moreover, recent experiments have demonstrated that the switching of a synthetic antiferromagnet does not obey the usual spin Hall angle-dependence, but that domain expansion and contraction can be selectively controlled toggling only the applied in-plane magnetic field magnitude and not its sign. We show for the first time that the combination of spin Hall torques and interlayer exchange coupling produces the necessary relative velocities for this switching to occur.
\end{abstract}

\maketitle

\section{Introduction}
Domain wall motion in ferromagnetic strips is a central theme in magnetization dynamics and has recently been instrumental to the discovery of several new current-induced effects.\cite{Miron2009,Miron2010,Miron2011,Emori2013,Ryu2013,Haazen2013} The attainable velocity of a domain wall driven by conventional spin-transfer torques (STTs)\cite{Zhang2004,Ralph2008,Brataas2012} is limited by the Walker breakdown,\cite{Schryer1974} upon which the domain wall deforms, resulting in a reduction of its velocity.

Current-induced torques derived from spin--orbit effects (SOTs) such as the spin Hall effect\cite{Ando2008,Emori2013,Ryu2013,Haazen2013} or an interfacial Rashba spin--orbit coupling\cite{Manchon2008,Kim2012a,Chernyshov2009} have enabled large domain wall velocities. We here consider the dependence of the domain wall velocity on the current and find that regardless of the relative importance of the reactive and dissipative components of the torque it is possible to achieve universal absence of Walker breakdown for all current densities for experimentally relevant spin--orbit coupling strengths. For spin--orbit torques other than the pure Rashba SOTs, such as the spin Hall torques, the velocity will not saturate as a function of current, but will increase linearly as long as a conventional spin-transfer torque is present. This behavior is robust against the presence of an interfacial Dzyaloshinskii--Moriya interaction\cite{Dzyaloshinskii1957,Moriya1960a,*Moriya1960,Fert1990} and is found both in perpendicular anisotropy ferromagnets, in shape anisotropy-dominated strips and in synthetic antiferromagnets (SAFs),\cite{Saarikoski2014,Lepadatu2016,Yang2015,Tomasello2016,Bi2017} where it enables very high domain wall velocites for relatively small current densities. Moreover, the combination of SOTs with the interlayer exchange torque was recently shown experimentally to produce novel switching behavior that circumvents the usual spin Hall angle-dependence.\cite{Bi2017} We show that the combination of spin Hall torques and interlayer exchange produces the required dependence of the domain wall velocity on the topological charge to qualitatively reproduce the experimental data.

\section{Universal Absence of Walker Breakdown}
\begin{figure}[b]
\includegraphics[width=.45\textwidth]{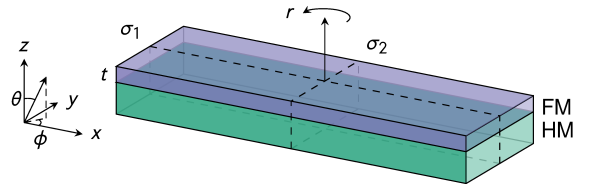}
\caption{\label{fig:system-single}Ultrathin ferromagnet with a heavy metal underlayer. We consider transverse domain wall motion along the $x$ axis. $r$, $\sigma_\text l$ and $\sigma_\text s$ denote the three nontrivial operations of the symmetry group $C_{2v}$.}
\end{figure}

We consider an ultrathin ferromagnet with a heavy metal underlayer as shown in \autoref{fig:system-single}. We describe the dynamics of the magnetization $\vec m(\vec r,t)$ using the Landau--Lifshitz--Gilbert (LLG) equation,\cite{Landau1935,*Landau2008,*Gilbert2004}
\begin{equation}\label{eq:llg}
\partial_t\vec m=\gamma\vec m\times\vec H-\frac{\alpha}{m}\vec m\times\partial_t\vec m+\vec\tau,
\end{equation}
where $\gamma<0$ is the gyromagnetic ratio, $m$ is the saturation magnetization, $\alpha<0$ is the Gilbert damping, $\vec H=-\delta F/\delta\vec m$ is the effective field acting on the magnetization and $\vec\tau$  is the current-induced torques. The free energy $F$ of the ferromagnet is a sum, 
\begin{equation}
F=\int\!\!\drm\vec r\,(f_\text Z+f_\text{ex}+f_\text{DM}+f_\text a),
\end{equation} 
of the Zeeman energy due to applied magnetic fields, the isotropic exchange, the interfacial Dzyaloshinskii--Moriya interaction and the magnetic anisotropy. 

The Zeeman energy and the isotropic exchange can be written respectively as $f_\text Z=-\vec H_0\cdot\vec m$, where $\vec H_0$ is the applied magnetic field, and $f_\text{ex}=(A/m^2)[(\nabla m_x)^2+(\nabla m_y)^2+(\nabla m_z)^2]$, where $A$ is the exchange stiffness.\cite{Gilbert2004} Inversion symmetry breaking at the interface between the heavy metal and the ferromagnet gives rise to an anisotropic contribution to the exchange known as the Dzyaloshinskii--Moriya interaction, which favors a canting of the spins.\cite{Dzyaloshinskii1957,Moriya1960a,*Moriya1960,Fert1990} The resulting contribution to the free energy is $f_\text{DM}=(D/m^2)[m_z(\nabla\cdot\vec m)-(\vec m\cdot\nabla)m_z]$, where $D$ is the magnitude of the Dzyaloshinskii--Moriya vector. Ultrathin magnetic films are prone to exhibit perpendicular magnetization due to interface contributions to the magnetic anisotropy.\cite{Handley2000} Consequently, we write the magnetic anisotropy energy as $f_\text a=-K_zm_z^2+K_ym_y^2$, corresponding to an easy axis in the $z$ direction and a hard axis in the $y$ direction. 

\subsection{Current-Induced Torques}
The current-induced torques $\vec\tau$ are conventionally divided into spin-transfer torques and spin--orbit torques. The spin-transfer torques can be written as\cite{Zhang2004,Ralph2008,Brataas2012} 
\begin{equation}
\vec\tau_\text{STT}=u\partial_x\vec m-\frac{\beta u}{m}\vec m\times\partial_x\vec m,
\end{equation} 
where $u=\mu_\text BPj/[em(1+\beta^2)]$ and $j$ is the electric current, $P$ is its spin polarization, $\mu_\text B$ is the Bohr magneton, $e$ is the electric charge and $\beta$ is the nonadiabacity parameter. The spin--orbit torques can be written as\cite{Manchon2008,Kim2012a,Chernyshov2009,Ando2008,Emori2013,Ryu2013,Haazen2013}
\begin{align}
&\vec\tau_\text R=\gamma\vec m\times H_\text R\vec e_y-\gamma\vec m\times\left(\vec m\times\frac{\beta H_\text R\vec e_y}{m}\right), \label{eq:rashba} \\
&\vec\tau_\text{SH}=\gamma\vec m\times\left(\vec m\times\frac{H_\text{SH}\vec e_y}{m}\right)+\gamma\vec m\times\beta_\text{SH}H_\text{SH}\vec e_y, \label{eq:spinhall}
\end{align}
where $H_\text R=\alpha_\text RPj/[2\mu_\text Bm(1+\beta^2)]$ and $\alpha_\text R$ is the Rashba parameter and where $H_\text{SH}=\hbar\theta_\text{SH}j/(2emt)$ and $\theta_\text{SH}$ is the spin Hall angle and $t$ is the magnet thickness. Since the spin Hall effect changes sign upon time-reversal, the principal spin Hall torque term is dissipative instead of reactive, in contrast to the principal term of the STTs and the Rashba SOTs.

In fact, assuming that the stack can be described using the $C_{2v}$ symmetry group (see \autoref{fig:system-single}) it can be shown that these torques exhaust the number of possible torque components. Hals and Brataas~\cite{Hals2013,*Hals2015} describe spin--orbit torques and generalized spin-transfer torques in terms of a tensor expansion. Assuming the lowest orders are sufficient to describe the essential dynamics, the reactive and dissipative spin--orbit torques are described by, respectively, an axial second-rank tensor and a polar third-rank tensor while the generalized spin-transfer torques are described using a polar fourth-rank tensor and an axial fifth-rank tensor. The torques that arise in a given structure are limited by the requirement that the tensors must be invariant under the symmetry operations fulfilled by the structure. We have assumed that the physical systems we consider are described by $C_{2v}$ symmetry. Combined with the fact that the current is applied in the $x$ direction only and that $\partial_y\vec m=0$ and $\partial_z\vec m=0$, this implies that there is only one relevant nonzero element in the axial second-rank tensor, two elements in the polar third-rank tensor, three elements in the polar fourth-rank tensor and six elements in the axial fifth-rank tensor.\cite{Birss1964}

The three relevant nonzero elements of the second- and third-rank tensors give rise to three spin--orbit torques. A detailed analysis shows that these torque components are captured by the Rashba and spin Hall torques in equations \eqref{eq:rashba} and \eqref{eq:spinhall}. As an aside, we note that although the Rashba and spin Hall effects may not necessarily capture all of the relevant microscopic physics~\cite{Kim2012b,Ryu2014,Fan2014} these torques can still be used to model the dynamics because they contain three \enquote*{free} parameters, $\alpha_\text R$, $\theta_\text{SH}$ and $\beta_\text{SH}$.

As has been shown in Ref.~\onlinecite{Hals2015}, the generalized spin-transfer torques reduce to the ordinary STTs in the nonrelativistic limit. Thus, by using the ordinary STTs we neglect possible spin--orbit coupling corrections to these higher-order terms.

\subsection{The Collective Coordinate Model}
\begin{figure}[tbp]
\includegraphics[width=.475\textwidth]{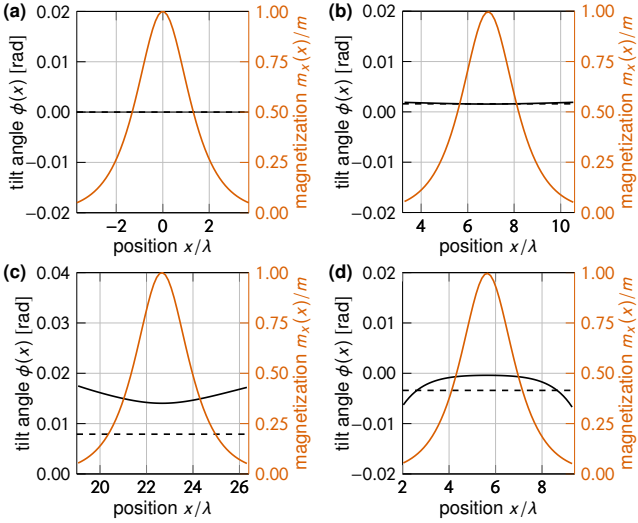}
\caption{\label{fig:phiplots} Position dependence of the domain wall tilt $\phi$. In each panel, the orange curve $m_x(x)$ shows the extension of the domain wall while the black solid curve shows the domain wall tilt $\phi(x)$ obtained by solving the full LLG equation \eqref{eq:llg} and the black dashed line shows the prediction of the collective coordinate model. (a)~Equilibrium solution. (b)~Spin-transfer torque dynamics. (c)~Spin Hall torque dynamics. (d)~Rashba spin--orbit torque dynamics. (a)--(d)~We use the material parameters supplied in the first column of \autoref{tab:par} with $j=\SI{5}{\mega\ampere\per\centi\meter\squared}$ except that $J=0$.}
\end{figure}

The magnetization is conveniently parametrized in spherical coordinates as $\vec m/m=\cos\phi\sin\theta\vec e_x+\sin\phi\sin\theta\vec e_y+\cos\theta\vec e_z$. Using the assumption that there is no magnetic texture along the $y$ and the $z$ axes, ${\nabla=\partial_x\vec e_x}$, we can find the domain wall profile by minimizing the free energy. The resulting Euler--Lagrange equations are
\[
A(\theta''\csc\theta\sec\theta-\phi^{\prime2})-D\phi'\sin\phi\tan\theta=(K_z+K_y\sin^2\phi)
\]
and
\[
A(\phi''+2\theta'\phi'\cot\theta)+D\theta'\sin\phi=K_y\cos\phi\sin\phi.
\]
One solution of these differential equations is the Néel wall solution $\phi=n\pi$ and ${\theta=}\,{2\arctan\exp[Q(x-X)/\lambda]}$, where $Q$ is the topological charge of the wall,\cite{Shibata2011} $X$ is the wall position and ${\lambda=\sqrt{A/K_z}}$ is the domain wall width. $n$ is even if ${D<0}$ and $Q=+1$, and $n$ is odd if ${D<0}$ and $Q=-1$. This domain wall profile is known as the Walker profile.\cite{Schryer1974} To be sure that $\phi=n\pi$ is really the global minimum, we solve the full LLG equation \eqref{eq:llg} for a single magnetic layer and let the solution relax without any applied currents or fields. The angle $\phi(x)$ can then be calculated as $\phi(x)=\arctan[m_y(x)/m_x(x)]$. However, $\phi(x)$ is ill defined in the domains where $\theta\to0\text{ or }\pi$. Consequently, we consider $\phi$ only inside the domain wall. As shown in \autoref{fig:phiplots}(a), the solution $\phi=0$ works very well.

Substitution of the Walker profile into the full LLG equation \eqref{eq:llg} using $\vec H_0=H_x\vec e_x$ and $Q=+1$ gives the collective coordinate equations, for the wall position $X$ and tilt $\phi$
\begin{align}
&\frac{\alpha\dot X}{\lambda}-\dot\phi=+\tfrac{\pi}{2}\gamma\Big(H_\text{SH}-\beta H_\text R\Big)\cos\phi+\frac{\beta u}{\lambda}, \label{eq:velocity-single} \\
&(1+\alpha^2)\dot\phi=-\frac{\alpha\gamma K_y}{m}\sin2\phi+\frac{\pi\alpha\gamma(D-H_xm\lambda)}{2m\lambda}\sin\phi \label{eq:tilt-single} \\
&-\frac{u(\alpha+\beta)}{\lambda}-\tfrac{\pi}{2}\gamma\Big[H_\text{SH}(1-\alpha\beta_\text{SH})-H_\text R(\alpha+\beta)\Big]\cos\phi. \notag
\end{align}
By doing this substitution, we are assuming that the domain wall moves as a rigid object described by two collective coordinates $X(t)$ and $\phi(t)$ (Ref.~\onlinecite{Shibata2011}). In particular, we are neglecting any position dependence in the domain wall tilt $\phi$. The collective coordinate model, or one-dimensional model, has been used previously to explain the qualitative behavior of both spin-transfer and spin--orbit torques.\cite{Emori2013,Ryu2013,Zhang2004,Schryer1974,Saarikoski2014,Lepadatu2016,Yang2015,Shibata2011,Beach2008,Ryu2014} However, it is important to remember that the model will always be an approximation, and we cannot necessarily expect quantitative agreement between experimental results and model predictions nor can we completely exclude the possibility of dynamics that is not captured by the one-dimensional model.\cite{Beach2008} We can nevertheless test the adequacy of the collective coordinate model by calculating $\phi(x)$ from a solution of the full LLG equation for a single magnetic layer, just as we did for the static case. As shown in \autoref{fig:phiplots}(b) the $x$ dependence of $\phi$ is negligible for spin-transfer torques. The $x$ dependence of $\phi$ is larger for spin Hall [\autoref{fig:phiplots}(c)] and Rashba spin--orbit torques [\autoref{fig:phiplots}(d)]. Nonetheless, the ability of the collective coordinate model to consistently qualitatively reproduce experimental behavior indicates that it captures the generality, if not all, of the physics in the system.

Equations \eqref{eq:velocity-single} and \eqref{eq:tilt-single} can be simplified by introducing $aj=\tfrac{\pi}{2}\gamma(H_\text{SH}-\beta H_\text R)$, $bj=\beta u/\lambda$, $c=-2\alpha\gamma K_y/m$, $d=\pi\alpha\gamma(D-H_xm\lambda)/(2m\lambda)$, $ej=-\tfrac{\pi}{2}\gamma[H_\text{SH}(1-\alpha\beta_\text{SH})-H_\text R(\alpha+\beta)]$ and $fj=-u(\alpha+\beta)/\lambda$. Walker breakdown is absent when the time derivative $\dot\phi$ vanishes, resulting in the condition
\begin{equation}\label{eq:walker}
0=c\sin\phi\cos\phi+d\sin\phi+j(e\cos\phi+f).
\end{equation}
Provided that the transverse domain wall is not transformed into for instance a vortex wall,\cite{Beach2008} Walker breakdown will be universally absent if $e>f$ because this equation always has a solution for $\phi$ regardless of the value of $j$. For increasing $j$, $\phi$ will level off to a value $\cos\phi=-f/e$. For realistic material values $e>f$ corresponds to a Rashba parameter $\alpha_\text R>4\mu_\text B^2/(\pi e\gamma\lambda)=\text{\num{1} to \SI{6}{\milli\electronvolt\nano\meter}}$ (pure Rashba SOTs) or a spin Hall angle $\theta_\text{SH}>4\mu_\text BPt/(\pi\hbar\gamma\lambda)=\text{\numrange{0,05}{0,09}}$ (pure spin Hall torques). To the best of our knowledge, the absence of Walker breakdown for spin Hall torques has not been noted previously, whereas absence of Walker breakdown for sufficiently strong Rashba spin--orbit coupling was pointed out in Ref.~\onlinecite{Linder2013}, and can also be noted in Refs~\onlinecite{Kim2012a} and~\onlinecite{Stier2014,Boulle2014,He2016}.

Let us write $\xi=\cos\phi$ and $\eta=\sin\phi$, so that $\xi^2+\eta^2=1$. Solving equation \eqref{eq:walker} for $\eta$ to get $\eta=-j(e\xi+f)/(c\xi+d)$, this relation gives a quartic equation
\[
c^2\xi^4+2cd\xi^3+[(ej)^2+d^2-c^2]\xi^2+2(efj^2-cd)\xi=d^2-(fj)^2.
\]
The exact solutions of the quartic are hopelessly complicated. However, they all have the same series expansion around $j=0$ and $j\to\infty$. We consider first the asymptotic expansion, 
\begin{equation}\label{eq:asymptotic}
\xi=-\frac{f}{e}+\frac{S_1}{j}+\mathcal O\!\left(j^{-2}\right),
\end{equation}
where $S_1$ represents the solutions of the quadratic equation $e^6\zeta^2=d^2e^4+c^2f^4+(c^2-d^2)f^2e^2+2cdef(f^2-e^2)$. Using equation \eqref{eq:velocity-single}, the wall velocity is then
\begin{equation}
\frac{\alpha\dot X}{\lambda}=\left(b-\frac{af}{e}\right)j+aS_1+a\mathcal O\!\left(j^{-1}\right).
\end{equation}
Back substitution of the abbreviations $a$, $b$, $e$ and $f$ shows that for pure Rashba SOTs the coefficient of the linear term reduces to zero because the ratio of the reactive to the dissipative torque is the same for the STTs and the Rashba SOTs. Thus, for large $j$ the domain wall velocity approaches a constant. For pure spin Hall torques we get instead the linear term $-u\alpha(1+\beta\beta_\text{SH})/[\lambda(1-\alpha\beta_\text{SH})]$. This means that \textit{for large $j$ the velocity is actually independent of the sign of the spin Hall angle and increases linearly with $j$.} Note the importance of including the STTs---which are always present---in these considerations: in the absence of STTs ($u\to0$) both $b$ and $f$ go to zero and the velocity levels off to a constant for large $j$ for any combination of SOTs. 

\begin{figure}[tbp]
\includegraphics[width=.45\textwidth]{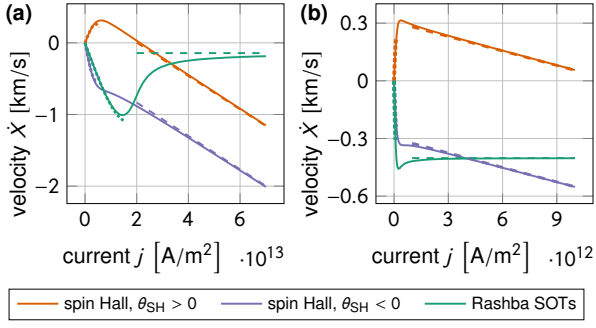}
\caption{\label{fig:numerics-single}Current--velocity relation for three different SOTs in the absence of Walker breakdown. The Rashba SOTs level off to a constant velocity at large currents, whereas the spin Hall torques asymptotically approach a linear current--velocity relation. Dashed lines show the asymptotic expansion and dotted curves show the series about $j=0$. We use the material parameters supplied in the (a)~first and (b)~second column of \autoref{tab:par} except that $J=0$.}
\end{figure}

For completeness, we also consider the series expansion about $j=0$, which gives 
\begin{equation}
\xi=-1+\frac{(e-f)^2}{2(c-d)^2}j^2+\mathcal O\!\left(j^4\right)
\end{equation}
and 
\begin{equation}
\frac{\alpha\dot X}{\lambda}=(b-a)j+\frac{a(e-f)^2}{2(c-d)^2}j^3+a\mathcal O\!\left(j^5\right).
\end{equation}
The key observation here is that in this regime the velocity does depend on the sign of the spin Hall angle ($a\propto\theta_\text{SH}$ for pure spin Hall torques) and increases with the cube of $j$. Combined with the spin Hall angle-independence of the velocity in the $j\to\infty$ limit, this implies that even in the absence of Walker breakdown a nonmonotonic current--velocity relation is possible. \autoref{fig:numerics-single}(a) shows a numerical solution of the coupled equations \eqref{eq:velocity-single} and \eqref{eq:tilt-single} as a function of $j$ for pure Rashba SOTs and for pure spin Hall torques both in the cases of $\theta_\text{SH}>0$ and $\theta_\text{SH}<0$ together with the analytical solutions close to $j=0$ and for large $j$ for parameters that are typical for a standard cobalt--nickel multilayer. We see that our analytical results successfully approximate the full solution in the expected ranges of validity indicating the absence of Walker breakdown in the numerical solution. 

The in-plane hard axis included in the magnetic anisotropy is appropriate for narrow ferromagnetic strips, which generally host Néel walls. Wider strips give Bloch walls,\cite{Handley2000} and by making the necessary modifications to the above calculations, we find that in this case the domain wall velocity retains the qualitative features elucidated above. This is also true for shape anisotropy-dominated strips, which host head-to-head walls. \textit{This shows that universal absence of Walker breakdown is a robust effect that does not depend on the details of the ferromagnetic material}, unlike other SOT effects studied previously.\cite{Khvalkovskiy2013} This fact is also illustrated by the numerics. In \autoref{fig:numerics-single}(b) we present numerical results obtained for a Néel wall in a PMA ferromagnet with anisotropies weaker by an order of magnitude, weaker magnetic damping and much larger Rashba spin--orbit coupling and spin Hall angle in the adjacent heavy metal. The results are qualitatively similar to those obtained in \autoref{fig:numerics-single}(a). 

\section{Coupled Domain Walls in a SAF Structure}
\begin{figure}[tbp]
\includegraphics[width=.45\textwidth]{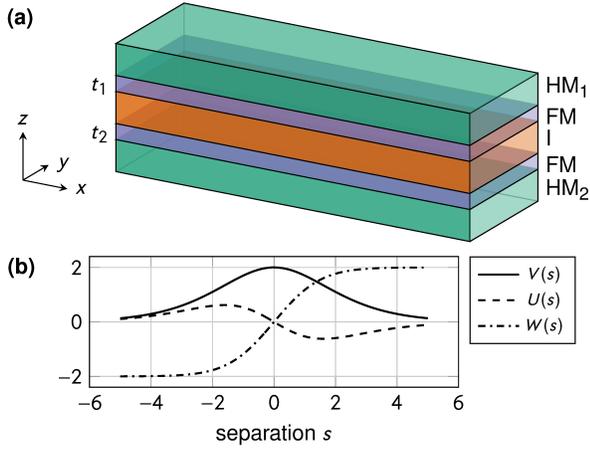}
\caption{\label{fig:system-saf}(a) Two ultrathin ferromagnets separated by an insulating spacer with heavy metal over- and underlayers. The ferromagnets are identical except for their thicknesses, but the different heavy metals induce different DMIs and SOTs. (b)~Dependence of the IEC terms $V(s)$, $U(s)$ and $W(s)$ on the wall separation.}
\end{figure}

We consider next an asymmetric stack of two ultrathin ferromagnets separated by an insulating spacer as shown in \autoref{fig:system-saf}(a). We describe the dynamics of each of the ferromagnets using separate LLG equations, but add to the free energy a coupling term,
\begin{equation}\label{eq:iec}
F_\text{IEC}=\!\!\int\!\!\frac{\drm\vec r_1}{m^{(1)}}\!\int\!\!\frac{\drm\vec r_2}{m^{(2)}}\,J(\vec r_1-\vec r_2)\left[\vec m^{(1)}(\vec r_1)\cdot\vec m^{(2)}(\vec r_2)\right],
\end{equation}
representing the interlayer exchange (IEC). We assume that the IEC is local in the plane, $J(\vec r_1-\vec r_2)=J\delta(x_1-x_2)\delta(y_1-y_2)$. Equation \eqref{eq:iec} then represent the lowest order coupling proposed by Bruno.\cite{Bruno1995} 

Following the same procedure as in the previous section we may now derive four coupled collective coordinate equations. With an antiferromagnetic coupling the walls will have opposite topological charges, $Q_2=-Q_1$. Since a local IEC can only affect the chiralities, and not the profiles of the walls, we can use the static solution derived previously, $\theta=2\arctan\exp[Q(x-X)/\lambda]$, where ${\lambda=\sqrt{A/K_z}}$ is the domain wall width and $Q$ is the topological charge. For a single wall the azimuthal angle $\phi$ is given by $\phi=n\pi$. $n$ is even if ${D<0}$ and $Q=+1$, and $n$ is odd if ${D<0}$ and $Q=-1$. To limit the scope of the treatment, we consider only the case where $D_1$ and $D_2$ have the same sign, $D_1<0$ and $D_2<0$. Then the DMI and the IEC cooperate to give the static solution $\phi_1=0$ ($Q_1=+1$) and $\phi_2=\pi$ ($Q_2=-1$). 

Substituting this static solution into the LLG equations using $\vec H_0=H_x\vec e_x$ gives the collective coordinate equations
\begin{widetext}
\begin{equation}
\begin{split}
(1+\alpha^2)\frac{\dot X_1}{\lambda}=&-\frac{\gamma K_y}{m}\sin2\phi_1+\frac{\pi\gamma(D_1-H_xm\lambda)}{2m\lambda}\sin\phi_1+\frac{\gamma Jt_2}{2m}\Big[\alpha U(s)\cos(\phi_1-\phi_2)+\alpha W(s)+V(s)\sin(\phi_1-\phi_2)\Big] \\
&-\frac{u(1-\alpha\beta)}{\lambda}+\tfrac{\pi}{2}\gamma\Big[H_\text{SH}^{(1)}\Big(\alpha+\beta_\text{SH}^{(1)}\Big)+H_\text R^{(1)}(1-\alpha\beta)\Big]\cos\phi_1, \label{eq:pos1}
\end{split}
\end{equation} 
\begin{equation}
\begin{split}
(1+\alpha^2)\frac{\dot X_2}{\lambda}=&+\frac{\gamma K_y}{m}\sin2\phi_2+\frac{\pi\gamma(D_2+H_xm\lambda)}{2m\lambda}\sin\phi_2-\frac{\gamma Jt_1}{2m}\Big[\alpha U(s)\cos(\phi_1-\phi_2)+\alpha W(s)-V(s)\sin(\phi_1-\phi_2)\Big] \\
&-\frac{u(1-\alpha\beta)}{\lambda}+\tfrac{\pi}{2}\gamma\Big[H_\text{SH}^{(2)}\Big(\alpha+\beta_\text{SH}^{(2)}\Big)+H_\text R^{(2)}(1-\alpha\beta)\Big]\cos\phi_2, \label{eq:pos2}
\end{split}
\end{equation}
\begin{equation}
\begin{split}
(1+\alpha^2)\dot\phi_1=&-\frac{\alpha\gamma K_y}{m}\sin2\phi_1+\frac{\pi\alpha\gamma(D_1-H_xm\lambda)}{2m\lambda}\sin\phi_1-\frac{\gamma Jt_2}{2m}\Big[U(s)\cos(\phi_1-\phi_2)+W(s)-\alpha V(s)\sin(\phi_1-\phi_2)\Big] \\
&-\frac{u(\alpha+\beta)}{\lambda}-\tfrac{\pi}{2}\alpha\gamma\Big[H_\text{SH}^{(1)}\Big(1-\alpha\beta_\text{SH}^{(1)}\Big)-H_\text R^{(1)}(\alpha+\beta)\Big]\cos\phi_1, \label{eq:angle1}
\end{split}
\end{equation}
\begin{equation}
\begin{split}
(1+\alpha^2)\dot\phi_2=&-\frac{\alpha\gamma K_y}{m}\sin2\phi_2-\frac{\pi\alpha\gamma(D_2+H_xm\lambda)}{2m\lambda}\sin\phi_2-\frac{\gamma Jt_1}{2m}\Big[U(s)\cos(\phi_1-\phi_2)+W(s)+\alpha V(s)\sin(\phi_1-\phi_2)\Big] \\
&+\frac{u(\alpha+\beta)}{\lambda}+\tfrac{\pi}{2}\alpha\gamma\Big[H_\text{SH}^{(2)}\Big(1-\alpha\beta_\text{SH}^{(2)}\Big)-H_\text R^{(2)}(\alpha+\beta)\Big]\cos\phi_2. \label{eq:angle2}
\end{split}
\end{equation}
\end{widetext}
where we have assumed that the bulk parameters of the two ferromagnets are equal and where $s$ is the separation between the two walls, $s=(X_1-X_2)/\lambda$. The IEC terms are expressed using the three functions $V(s)$, $U(s)$ and $W(s)$; 
\begin{align*}
&V(s)=2s\csch s, \\
&U(s)=2\csch s-2s\coth s\csch s, \\
&W(s)=2\coth s-2s\csch^2s.
\end{align*}
These functions are plotted in \autoref{fig:system-saf}(b). 

\begin{figure}[tbp]
\includegraphics[width=.45\textwidth]{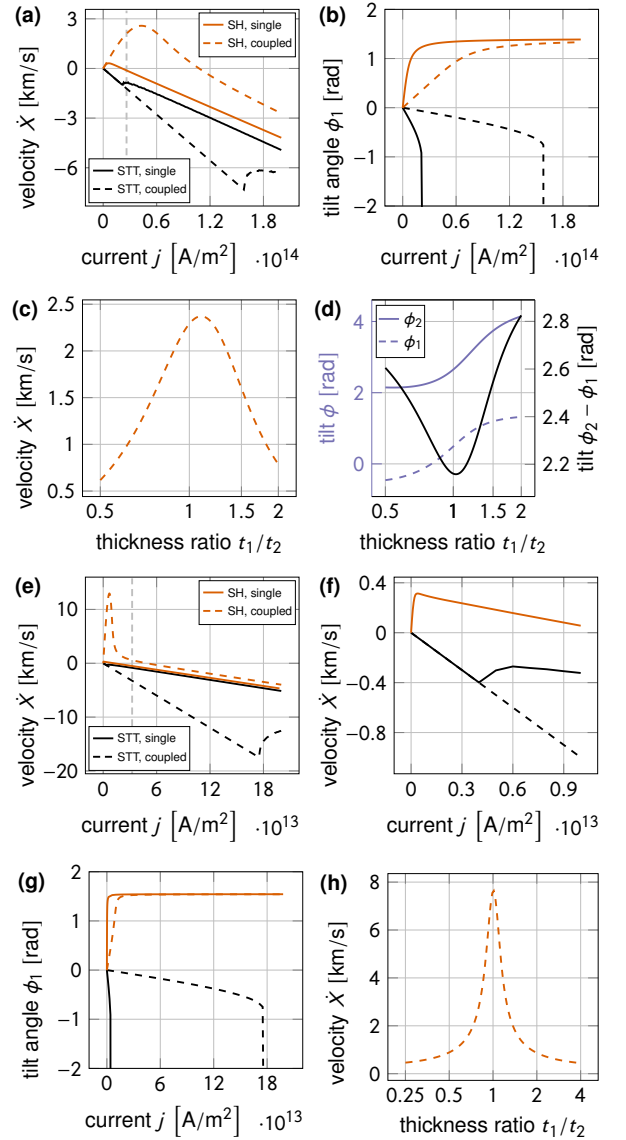}
\caption{\label{fig:numerics-saf}Domain wall dynamics in interlayer exchange coupled ferromagnets. (a)~and (b)~The IEC delays Walker breakdown for STT driving, but the subcritical differential velocity remains unaffected. With spin Hall torques the tilt angle stabilizes at a finite value, indicating universal absence of Walker breakdown. The tilt angle approaches its limiting value more slowly in the presence of IEC. (c)~and (d)~The IEC gives the velocity a nonmonotonic thickness-dependence resulting in a peak close to $t_1/t_2=1$. [$j=\SI{3}{\giga\ampere\per\centi\meter\squared}$, corresponding to the dashed vertical line in~(a).] We use the material parameters supplied in the first column of \autoref{tab:par}. (e)--(h)~These results are robust against a change in parameters to those in the second column of \autoref{tab:par}.}
\end{figure}

Equations \eqref{eq:pos1} and \eqref{eq:angle1} reduce to equations \eqref{eq:velocity-single} and \eqref{eq:tilt-single} when $J\to0$. To solve equations \eqref{eq:pos1}--\eqref{eq:angle2} numerically, we rescale the equations to obtain dimensionless variables. The dimension of equations \eqref{eq:pos1}--\eqref{eq:angle2} is \si{\hertz}. A convenient scaling factor with the same dimensions is $\mu_0\gamma m$. By dividing equations \eqref{eq:pos1}--\eqref{eq:angle2} by $\mu_0\gamma m$ we get the rescaled variables $\tilde t=t\mu_0\gamma m$, $\tilde X_i=X_i/\lambda$, $\tilde H_x=H_x/\mu_0m$, $\tilde K_y=K_y/\mu_0m^2$, $\tilde D_i=D_i/\mu_0m^2\lambda$, $\tilde t_i=t_i/\lambda$, $\tilde J=J\lambda/\mu_0m^2$ and $\tilde u=u/\mu_0\gamma m\lambda$. We solve the equations using an explicit fourth order Runge--Kutta scheme with adaptive stepsize control, implemented as a Dormand--Prince pair.\cite{Dormand1980}

\subsection{Universal Absence of Walker Breakdown in SAF structures}
For parameter values representative of a standard cobalt--nickel multilayer we obtain the current--velocity and current--tilt relations shown in \autoref{fig:numerics-saf}(a) and (b) for $t_1/t_2=1$ in the case where only STTs are present and in the case where spin Hall torques are additionally present. We see that the presence of the IEC delays Walker breakdown when the wall is driven by ordinary STTs, but the subcritical differential velocity remains unaffected. This can also be shown analytically by solving for the tilt angle of the wall as a function of current. Such a calculation shows that the tilt angle is suppressed by the IEC (but the breakdown angle is still $\pi/4$). Back-substitution of this angle into the torque acting on the wall shows that this torque is independent of $J$, explaining why there is no change in the differential velocity. 

When spin Hall torques are included, the domain wall tilt levels off to a finite value and the current--velocity relation is linear in the $j\to\infty$ limit. \textit{This shows that universal absence of Walker breakdown is also found in SAF structures.} The effect of the IEC can be understood simply as a rescaling of the constant $S_1$ and the higher order constants $S_2,S_3,\dots$ in the expansion \eqref{eq:asymptotic}, making the tilt angle approach its limiting value more slowly. Thus, the effect of the IEC on both the STT and spin Hall results is to suppress the domain wall tilt, as shown in \autoref{fig:numerics-saf}(b). We note that the combination of spin Hall torques and IEC produces much higher domain wall velocities than in single ferromagnets for comparatively small current densities.\cite{Yang2015}

In a single ferromagnet the velocity of a wall driven by spin Hall torques decreases with $t$ as $1/t$. When changing $t_2$ from $t_2=t_1/2$ to $t_2=2t_1$ in a SAF structure, we find that the velocity peaks close to $t_1/t_2\approx1$, which maximizes the IEC torque [see \autoref{fig:numerics-saf}(c); the deviation from 1 is due to the DMI]. This can be understood by considering \autoref{fig:numerics-saf}(d); at $t_1/t_2\approx1$ the magnetizations in both layers are tilted in the $y$ direction. Increasing (decreasing) $t_2$ to $t_2=2t_1$ ($t_2=t_1/2$) reduces (increases) $H_\text{SH}^{(2)}$ and increases (reduces) $H_\text{IEC}^{(1)}$, thus $(\phi_2-\phi_1)$ approaches $\pi$ and the IEC torque is reduced. 

Just as for the single ferromagnetic layer the results for the coupled walls are robust against a change of parameters, as shown in \autoref{fig:numerics-saf}(e)--(h). 

\begin{table}[tbp]
\caption{\label{tab:par}Parameters used for the numerical solution of equations \eqref{eq:pos1}--\eqref{eq:angle2} and for analytical estimates in the text.}
\begin{tabular}{lSSSs}
\toprule
parameter & {Co--Ni} & {strong SOC} & {Bi \textit{et al.}} & {unit} \\
\midrule
gyromagnetic ratio $\gamma$ & -0,19 & -0,19 & -0,19 & \tera\hertz\per\tesla \\
domain wall width $\lambda$ & 4 & 16 & 2 & \nano\meter \\ 
hard axis anisotropy $K_y$ & 200 & 20 & 2 & \kilo\joule\per\meter\cubed \\
saturation magn. $m$ & 1 & 1 & 1,1 & \mega\ampere\per\meter \\
DM constant $D$ & -1,4 & -1,0 & -0,1 & \milli\joule\per\meter\squared \\
Gilbert damping $\alpha$ & -0,25 & -0,1 & -0,5 & \\
spin-polarization $P$ & 0,5 & 0,5 & 0,5 & \\
nonadiabacity param. $\beta$ & 0,5 & 0,4 & 2 & \\
Rashba parameter $\alpha_\text R$ & 6,3 & 75 & & \milli\electronvolt\nano\meter \\
spin Hall angle $\theta_\text{SH}$ & 0,1 & 0,2 & 0,12 & \\
spin Hall $\beta$-term $\beta_\text{SH}$ & 0,02 & 0,02 & 0,02 & \\
interlayer exchange $Jt_1t_2$ & 5 & 5 & 1,5 & \milli\joule\per\meter\squared \\
thickness $t_1$ & 1,2 & 1,2 & 0,6 & \nano\meter \\
thickness $t_2$ & 1,2 & 1,2 & 1,7 & \nano\meter \\
\bottomrule
\end{tabular}
\end{table}

\subsection{Novel Switching Behavior in SAF Structures}
\begin{figure}[t]
\includegraphics[width=.45\textwidth]{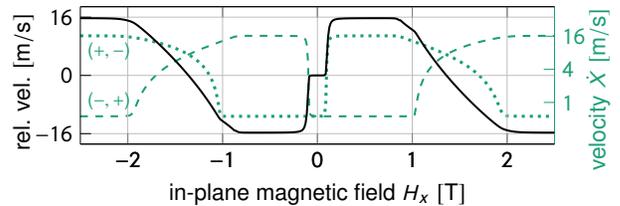}
\caption{\label{fig:cbi}Qualitative reproduction of the experimental results of Bi \textit{et al.}\cite{Bi2017} The sign of the relative velocity of walls with $(Q_1,Q_2)=(+1,-1)$ and $(Q_1,Q_2)=(-1,+1)$ can be toggled only by changing the magnitude of the applied field. We use the material parameters supplied in the third column of \autoref{tab:par}.}
\end{figure}

\citet{Bi2017} have very recently demonstrated completely novel switching behavior in SAF structures. In single ferromagnets, domain walls with one topological charge will travel faster than those with the opposite topological charge if an in-plane magnetic field is applied.\cite{Je2013} If the relative velocity is large enough the favored domains can overcome the destabilizing action of the current (see Refs~\onlinecite{Shibata2005,Nakatani2008,Torrejon2015,Taniguchi2015}) and merge.\cite{Yu2014,Garello2014,Rojas-Sanchez2016} The favored magnetization direction is uniquely determined by the spin Hall angle and the applied magnetic field for a fixed direction of the current. \citeauthor{Bi2017} observed this behavior in SAF structures for small in-plane fields, but by toggling between large and small values of the in-plane field (same sign), they were able to toggle the sign of the relative velocity of the walls and thereby the favored magnetization direction. Using material parameters that approximate the samples of \citeauthor{Bi2017}, our model is the first to qualitatively reproduce this behavior, as shown in \autoref{fig:cbi}. Under an in-plane field in the range \SIrange{0,3}{1,4}{\tesla}, walls with $(Q_1,Q_2)=(+1,-1)$ travel faster than walls with $(Q_1,Q_2)=(-1,+1)$ and `up' magnetization is favored. If the field is increased beyond \SI{1,4}{\tesla},  the relative velocity changes sign, and `down' magnetization is favored. (The offset from zero is due to the DMI.)

\section{Conclusion}
We have shown that complete suppression of Walker breakdown is possible in a wide range of domain wall systems driven by spin--orbit torques, including head-to-head walls in soft magnets, Bloch and Néel walls in perpendicular anisotropy magnets, in the presence of the Dzyaloshinskii--Moriya interaction and in coupled domain walls in synthetic antiferromagnets. For spin--orbit torques other than pure Rashba spin--orbit torques this leads to a linear current--velocity relation instead of a saturation of the velocity for large currents. In combination with interlayer exchange coupling, spin--orbit torque driven domain wall motion in synthetic antiferromagnets gives rise to novel switching behavior and very high domain wall velocities.

\begin{acknowledgments}
Funding via the \enquote{Outstanding Academic Fellows} program at NTNU, the COST Action MP-1201, the NV Faculty, and the Research Council of Norway Grants No.~216700 and No.~240806, is gratefully acknowledged. We thank Morten Amundsen for very useful discussions of the numerics. 
\end{acknowledgments}

\newpage

\end{document}